\documentclass[12pt]{article}
\usepackage{geometry}
\usepackage{a4}
\usepackage{graphicx}
\usepackage{mathtools}
\usepackage{epsf}
\usepackage{amsmath}
\usepackage{amssymb}
\usepackage{braket}
\usepackage{cite}
\newcommand{\be}{\begin{equation}}
\newcommand{\ee}{\end{equation}}

\newcommand{\Rmnum}[1]{\expandafter\@slowromancap\romannumeral #1@}
\newcommand{\bea}{\begin{eqnarray}}
\newcommand{\eea}{\end{eqnarray}}

\begin{document}
\def\A{{\mathbb{A}}}
\def\B{{\mathbb{B}}}
\def\C{{\mathbb{C}}}
\def\P{{\mathbb{P}}}
\def\R{{\mathbb{R}}}
\def\s{{\mathbb{S}}}
\def\T{{\mathbb{T}}}
\def\Z{{\mathbb{Z}}}
\def\W{{\mathbb{W}}}
\begin{titlepage}
\title{Information Geometry in Time Dependent Quantum Systems and the Geometric Phase}
\author{}
\date{
Anshuman Dey, Suvankar Paul, Pratim Roy, Tapobrata Sarkar
\thanks{\noindent E-mail:~ deyanshu, svnkr, proy, tapo @iitk.ac.in}
\vskip0.4cm
{\sl Department of Physics, \\
Indian Institute of Technology,\\
Kanpur 208016, \\
India}}
\maketitle
\abstract{
\noindent
We study information theoretic geometry in time dependent quantum mechanical systems.  First, we discuss global properties of the parameter manifold for 
two level systems exemplified by i) Rabi oscillations and ii) quenching dynamics of the XY spin chain in a 
transverse magnetic field, when driven across anisotropic criticality. 
Next, we comment upon the nature of the geometric phase from classical holonomy analyses of such parameter manifolds. In the context of the transverse XY model
in the thermodynamic limit, our results are in contradiction to those in the existing literature, and we argue why the issue deserves a more careful analysis. 
Finally, we speculate on a novel geometric phase in the model, when driven across a quantum critical line. 
}
\end{titlepage}
\section{Introduction}
Geometric notions often provide new insights into physical systems - one of the most profound examples being the adiabatic Berry phase \cite{Berry} or 
the more general geometric phase \cite{AA1},\cite{AA2}, in time dependent quantum systems. 
These phases, which are experimentally measurable, arise over and above the dynamical phase in 
quantum mechanics, due to the fact that the parameter space (typically, the space of coupling constants) of a quantum system may be curved, and can be attributed 
to the holonomy arising out of such a curvature. 

Ever since the discovery of the Berry phase more than three decades back, a lot of attention has been paid to such geometrical aspects in physics and the literature on the
subject is, by now, vast. As is standard in the literature, this will be refereed to as information theoretic geometry (IG) in sequel, and the space of parameters
as the parameter manifold (PM). Typical studies of IG consist of constructing a Riemannian metric on the parameter manifold \cite{pv}. As is also well known, this is
a pullback of the Fubini-Study metric on a projective Hilbert space. 

In this paper, we focus on two aspects of the IG of time dependent quantum mechanical systems.\footnote{Serval aspects of IG in the context of many body
quantum systems have been studied by us earlier, see, e.g \cite{tapo}.} First, we study the global nature of the parameter
manifold in such time dependent situations. Here, we mostly consider two level systems, and consider a PM that consists of time and one coupling constant. IG for these
systems is expected to be spherical, as the geometry is captured by the Bloch sphere. This is indeed the case for single particle systems, but we show that
for many body systems that undergo phase transitions, the situation is different. In particular, we consider the transverse XY spin chain, undergoing
an anisotropic quantum quench, and show that the parameter manifold is spherical everywhere except at critical points. 

The second aspect of IG that we study in this paper concerns the holonomy of the PM for many body quantum systems. In particular, we ask
what the ``classical holonomy'' (as measured by a classical spin) of the PM teaches us about the Berry phase or geometric phase of such a system. 
In simple cases (for example in spin-$j$ systems in a rotating magnetic field) this is well known. Here, we ask the 
same question for a many body quantum systems, in particular the transverse XY spin chain. We find that for this example, the two exhibit similar features and 
captures the same information about quantum phase transitions. 

Importantly, contrasting the classical holonomy of the transverse XY spin chain with its quantum counterpart in the thermodynamic limit, 
we encounter a contradiction with results in the existing literature \cite{Zhu}, and point out its possible resolution. Our computations here point to the fact that the 
Berry phase for this model deserves more careful analysis. Finally, in the context of the same model, we speculate on the possibility of a novel 
geometric phase from a purely classical analysis. 

In the rest of the paper, we elaborate upon the points mentioned above. 

\section{Information Geometry and Time Dependent Quantum Systems}

We first briefly review the results of \cite{pv}, where the metric on the parameter manifold of quantum systems was constructed.  This is the construction 
of a quantum geometric tensor, whose real part is
the Riemann metric on the PM. The main idea is to consider two quantum states, infinitesimally separated in the parameters, and construct the quantity
\begin{equation}
|\psi\left(q +dq\right) - \psi\left(q\right)|^2 = \langle \partial_i \psi|\partial_j \psi\rangle dq^idq^j = \alpha_{ij}dq^idq^j~,
\label{pv1}
\end{equation}
where $q^i$ (collectively denoted as $q$) are the parameters (or coupling constants), and
$\partial_i \equiv \partial/\partial q^i$. It can be checked that $\alpha_{ij}$s are not gauge invariant. In order to overcome this, another second order
tensor is introduced : 
\begin{equation}
g_{ij} = \alpha_{ij} - \beta_i\beta_j;~~~~\beta_i = -i\langle\psi\left(q\right)|\partial_i\psi\left(q\right)\rangle~.
\label{metric}
\end{equation}
This is a meaningful gauge invariant quantity, and the real part of $g_{ij}$ specify a metric on the parameter manifold, which is induced from the 
structure of the Hilbert space. Following \cite{ZanardiPRL}, we will briefly elaborate on this, for completeness. 

To describe IG, given a (complex) Hilbert space ${\mathcal H}$ of normalised vectors, one constructs a projective Hilbert space ${\mathcal P}({\mathcal H})$ which is a manifold 
of rays, i.e vectors in ${\mathcal H}$ that differ by a phase factor are considered equivalent. ${\mathcal P}({\mathcal H})$ admits a complex valued 
quantum geometric tensor, whose real part is the Riemannian metric on ${\mathcal P}({\mathcal H})$, and the metric on the PM of eq.(\ref{pv1}) is its pull back.
Mathematically, this can be seen as follows.  If we consider two infinitesimally separated (pure) states $|\psi\rangle$ and $|\psi + \delta \psi\rangle$ 
in the projective Hilbert space, the Fubini-Study distance is  $ds = \cos^{-1}|\langle\psi|\psi+\delta \psi\rangle|$. This can be 
expanded as $ds^2 = 2(1- |\langle\psi|\psi+\delta \psi\rangle|$), and upon Taylor expanding $|\psi + \delta \psi\rangle$, and using the pull back
$\delta|\psi\rangle =  \partial_{i}|\psi\rangle dq^i$, one obtains the metric of eq.(\ref{pv1}). 
This metric is positive definite and gives rise to a Riemannian structure via a line element $ds^2 = \sum_{i,j} g_{ij}dq^idq^j$. 
Note that if time is taken to be a parameter of the system, the definition of energy operator\footnote{We will set $\hbar = 1$ 
throughout this paper.} $E = i\partial/\partial t$ implies from eq.(\ref{pv1}) and
eq.(\ref{metric}) that $g_{tt} = (\Delta E)^2$, where $\Delta E$ is the energy uncertainty. This is of course the same result derived in \cite{AA2}.

\subsection{Information Geometry and Two Level Systems}

We now turn to two-level systems that will be our main interest here. In the interaction picture, a general state ket is represented as 
\begin{equation}
|\psi(t)\rangle = c_1(t)|e_1\rangle + c_2(t)|e_2\rangle~,
\label{genket}
\end{equation}
where $|e_1\rangle$ and $|e_2\rangle$ are the base kets for the time-independent problem, and the time dependent coefficients $c_1(t)$ and $c_2(t)$ can
also be functions of the other coupling constants of the theory. Then, if there is one such coupling constant apart from time, 
using eq.(\ref{genket}) in eq.(\ref{metric}), we arrive at the components of the geometric tensor on the PM :
\begin{eqnarray}
g_{00} &=& {\dot c_1}{\dot c_1}^* + {\dot c_2}{\dot c_2}^* + \left({\dot c_1}c_1^* + {\dot c_2}c_2^*\right)^2 ~,\nonumber\\
g_{01} &=& {\dot c_1}^*c_1' + {\dot c_2}^*c_2' + \left(c_1^*{\dot c_1} + c_2^*{\dot c_2}\right)\left(c_1^*c_1' + c_2^*c_2' \right) = g_{10} ~,\nonumber\\
g_{11} &=& c_1'c_1^{*'} + c_2'c_2^{*'} + \left(c_1'c_1^* + c_2'c_2^*\right)^2~.
\label{tdmetric}
\end{eqnarray}
Here, to make the notation uniform, we have used the symbol $0$ for time and $1$ for the other coordinate (coupling constant) and the dot and prime 
denote derivatives with respect to these, respectively. It is also to be noted that the real parts of the components of the tensor of eq.(\ref{tdmetric}) is
the metric tensor. 

For two level systems, IG might seem somewhat obvious. This is roughly because the geometry of the space is the Bloch sphere, which for the two
dimensional Hilbert space that we have in mind, is simply the complex projective space $\C\P^1$. The metric of eq.(\ref{tdmetric}), should
thus be simply the spherical metric (in a possibly complicated set of coordinates). This is indeed true, except for systems that exhibit phase transitions.  
Our purpose here would be to contrast features of such systems with simpler ones. We therefore recapitulate a few simple situations, before
we study a many body system in the next subsection. 

To set the stage, let us consider non-adiabatic evolution of a spin $1/2$ particle in a magnetic field with frequency $\omega$. We refer the reader to \cite{Wang} for details,
where this system was first solved, and it was shown that in an appropriate body-fixed coordinate system, for cyclic non-adiabatic evolution, the energy 
eigenkets for spin $j$ are given by 
\begin{equation}
|\psi_m\rangle = {\rm e}^{-iE_mt}\sum_n d^j_{nm}(\alpha){\rm e}^{-in\omega t}|j,n\rangle ~,
\label{Wangeq}
\end{equation}
where $d^j_{nm}(\alpha)$ denotes the Wigner $d$-function of rank $j$, $E_m$ is the energy eigenvalue for the $m$ th state in the time-independent problem, 
and $m$ can take on any of the $2j+1$ values. Here, $\alpha$ is related to the the polar 
angle $\theta$ (the direction of the magnetic field) by a complicated formula (equations (11.c) and (11.d) of \cite{Wang}) which we do not reproduce here for brevity. 
It is easy to check that for spin half particles, eq.(\ref{metric}) gives a spherical metric, $ds^2 = (1/4)(d\alpha^2 + \omega^2 \sin^2\alpha dt^2)$, with the identification
$\phi = \omega t$. 

Now we consider a two-state system with the Hamiltonian (with real $\gamma$)
\begin{equation}
H = E_1|e_1\rangle\langle e_1| + E_2 |e_2\rangle\langle e_2| + \gamma {\rm e}^{i\omega t}|e_1\rangle\langle e_2| + h.c ~.
\end{equation}
This is the Rabi problem, and is exactly solvable \cite{Sakurai}. One obtains, with boundary condition $c_1(0)=1$, i.e with the system being  
initially in the sate $|e_1\rangle$, and denoting $\Delta = \omega - (E_2-E_1)$ (with $E_2 > E_1$), $|\psi(t)\rangle = c_1(t)|e_1\rangle + c_2(t)|e_2\rangle$ where
\begin{equation}
c_1(t)=e^{\frac{i \Delta  t}{2}} \left[\cos \left(\frac{t}{2} \sqrt{\alpha}\right)-\frac{i \Delta}{\sqrt{\alpha}}  \sin \left(\frac{t}{2}\sqrt{\alpha}\right)\right] ~,
~c_2(t)=-\frac{2i \gamma}{\sqrt{\alpha}}  e^{-\frac{1}{2} i \Delta  t} \sin \left(\frac{t}{2}\sqrt{\alpha}\right)~.
\label{probabs}
\end{equation}
Here we have denoted $\alpha = 4\gamma^2 + \Delta^2$. 
We can now work out the IG of the system, by putting the above solution in eq.(\ref{tdmetric}). The two coordinates are taken to 
be $t$ and $\gamma$, which are denoted by $0$ and $1$ respectively, following our convention outlined before. $\Delta$ is taken to be fixed. 
We list the metric components below :
\begin{eqnarray}
g_{00}&=&\frac{\gamma ^2}{\alpha^2} \left[16 \gamma ^4+2 \gamma ^2 \Delta ^2+\Delta ^4-2\gamma ^2 \Delta ^2 \left(\cos 2\tau-4
\cos \tau\right)\right]~,\nonumber\\
g_{01}&=&\frac{1}{\alpha^3}\left[16 \alpha  \gamma ^5 t+\gamma  \Delta ^2 \left(\sqrt{\alpha }\sin \tau \left(8 \gamma ^2+\Delta ^2-4
\gamma ^2 \cos \tau\right)+4 \alpha \gamma ^2 t \cos \tau\right)\right]~,\nonumber\\
g_{11}&=&\frac{2}{\alpha^3} \left[3 \gamma ^2 \Delta ^2+\Delta ^4+8 \tau^2  \gamma ^4 +\Delta ^2 \left(\gamma ^2 \left(4 \tau \sin
\tau+\cos 2\tau\right)-\alpha  \cos \tau\right)\right]~,
\label{Rabimetric}
\end{eqnarray}
where we have defined $\tau \equiv t\sqrt{\alpha}$. It can be checked that 
$g_{00}$ has oscillatory behaviour, with its maxima and minima coinciding with those of the probabilities obtained from eq.(\ref{probabs}). 
The scalar curvature of the PM can be computed from a standard formula in differential geometry \footnote{For a two dimensional manifold with 
coordinates $(x,y)$, the scalar curvature is ($g$ is the determinant of the metric tensor) 
$$R = \frac{1}{g}\left[\partial_x\left(\frac{g_{xy}}{g_{xx}g}\partial_y g_{xx} - \frac{1}{g}\partial_xg_{yy}\right)
+ \partial_y\left(\frac{2}{g}\partial_xg_{xy} - \frac{1}{g}\partial_yg_{xx} - \frac{g_{xy}}{g_{xx}g}\partial_xg_{xx}\right)\right].$$ In the sign convention used 
here, the curvature of the two sphere is positive.} and is expectedly 
a positive constant (with our units, $R=8$). This indicates that the parameter manifold is spherical, as before. Of course, the complicated nature
of the metric of eq.(\ref{Rabimetric}) makes it difficult to cast it into a manifestly spherical form. This is however not a problem, since the scalar curvature is 
invariant under coordinate transformations.

Let us point out a few important features that we will use in our analysis of the next subsection. First, we consider the limit $\Delta \to 0$, i.e the system is at resonance. 
In this limit, from eq.(\ref{Rabimetric}), we find that $g_{00}=\gamma^2$, $g_{01}=t\gamma$, $g_{11} = t^2$. It might thus seem that the information metric becomes undefined
in this limit, since the determinant of the metric vanishes. However, a careful calculation of the scalar curvature, where the limit is taken at the end,  
reveals that this is still $R=8$. Next, we consider the limit $\gamma \to 0$. This case is the limit of no interaction, i.e the system remains in the initial state 
at $t=0$. We expect that $g_{00}$ should be zero in this case, as the system is in an energy eigenstate and $\Delta E = 0$. One can check that this is 
indeed the case, and that $g_{01}$ equals zero as well in this limit, while $g_{11} = (2/\Delta^2)(1-\cos(t\Delta))$. However, the scalar curvature is $R=8$, when the limit is 
taken at the end. 

It is also interesting to compute the geometric phase for the system when it goes through one full cycle. We will review this in some details in
the beginning of the next section. For the time being, let us note that in our case, the geometric phase can be obtained
as $\beta = -i\int_0^{T} \langle \psi(t)|\partial_0 |\psi(t)\rangle dt$ where $T = 4\pi/\sqrt{\alpha}$, and a simple calculation reveals 
\begin{equation}
\beta = -\frac{8 \pi  \gamma ^2 \Delta }{\left(4 \gamma ^2+\Delta ^2\right)^{3/2}}~.
\end{equation}

\subsection{Information Geometry and a Quantum Quench}

We will now move on to study the more interesting example of the IG of a many body system, the transverse XY spin chain. Various aspects of this model have been studied 
in details over the last many decades, starting from \cite{LSM}, and we will only recapitulate the salient features that are required for our analysis. 
The model is described by the Hamiltonian
\begin{equation}
H = -\frac{1}{2}\sum_n \left(J_x \sigma^x_i\sigma^x_{i+1} + J_y \sigma^y_i\sigma^y_{i+1} + h\sigma^z_i \right)~.
\label{xyHamil}
\end{equation}
Here,  $J_x$ and $J_y$ measure the anisotropy in the $x$ and $y$ directions, ${\vec\sigma}$ denotes the Pauli spin matrices, and $h$ is
the transverse magnetic field. The Hamiltonian can be diagonalised by a series of Jordan-Wigner, Fourier and Bogoliubov transformations, and decouples 
in the momentum representation into as a sum $H = \sum_{\oplus k>0}H_k$, where $k$ denotes the $k$'th momentum mode, and 
$H_k$ operates on a four-dimensional Hilbert space spanned by $|0\rangle$, $|\pm k\rangle$ and $|k,-k\rangle$. The gap in the energy spectrum is given by 
\begin{equation}
\epsilon_k = \left[\left(h+\cos k \left(J_x+J_y\right)\right){}^2+\sin ^2 k\left(J_x-J_y\right){}^2\right]^{1/2}~.
\label{gap}
\end{equation}
The energy gap closes at $h = \mp(J_x + J_y)$ for $k=0, \pi$, signalling a quantum phase transition from a paramagnetic to a ferromagnetically ordered phase, 
dubbed as the Ising transition in the literature. Further, for $J_x=J_y$, the energy gap closes for $k = \cos^{-1}(-h/2J_x)$, and this is known as the anisotropic transition. 

Our interest would be in understanding the IG of this model when it undergoes a quench i.e, is driven across a critical point. Quenching dynamics was studied in this model 
for transverse quench in \cite{Cherng} and for anisotropic quench in \cite{Amit}. The essential idea here is 
that since $|\pm k\rangle$ are eigenstates of $H_k$, one makes $H_k$ a function of time (by appropriately making some of the coupling constants linearly time dependent), 
and projects it onto the subspace spanned by $|0\rangle$, $|k,-k\rangle$. The resulting Schrodinger equation is then seen to be identical to a Landau-Zener problem 
with an avoided crossing \cite{Cherng},\cite{Amit}. 

Here we will focus on the anisotropic quenching scheme of \cite{Amit}, and set $J_x=t/\tau$, with $\tau$ being a time scale, which we will set to unity for numerical convenience. 
From our discussion earlier, this choice of $J_x$ corresponds to driving the system through the anisotropic transition line $J_x=J_y$. In order to compute the information 
metric at late times, one has to essentially solve a time dependent two-state problem in the basis $|e_{ik}\rangle$, $i=1,2$ where the basis vectors are suitable linear combinations
of the vectors $|0\rangle$ and $|k,-k\rangle$. Writing a general state vector $|\psi_k(t)\rangle = c_1(t)|e_{1k}\rangle + c_2(t)|e_{2k}\rangle$, the resulting Schrodinger 
equation (eq.(12) of \cite{Amit}), can be solved in terms of Weber functions (see section titled ``Landau Zener Theory.''  in \cite{Suzuki}) for late times. The solutions can then be used in 
eq.(\ref{tdmetric}) to compute the metric at such times, with coordinates $(t, J_y)$. The expressions for the metric, which involve the derivatives of Weber functions 
are very cumbersome and will not attempt to reproduce them here. Rather, we focus on some salient physical features of our results via a graphical analysis. Throughout the following, 
we will set the transverse field $h=0.5$. Importantly, we remember that we are not taking the strict $t\to \infty$ limit as is usual in Landau-Zener problems. Rather, we
use the asymptotic expansions of Weber functions \cite{Suzuki} for large times. 

\begin{figure}[t!]
\begin{minipage}[b]{0.5\linewidth}
\centering
\includegraphics[width=3.0in,height=2.3in]{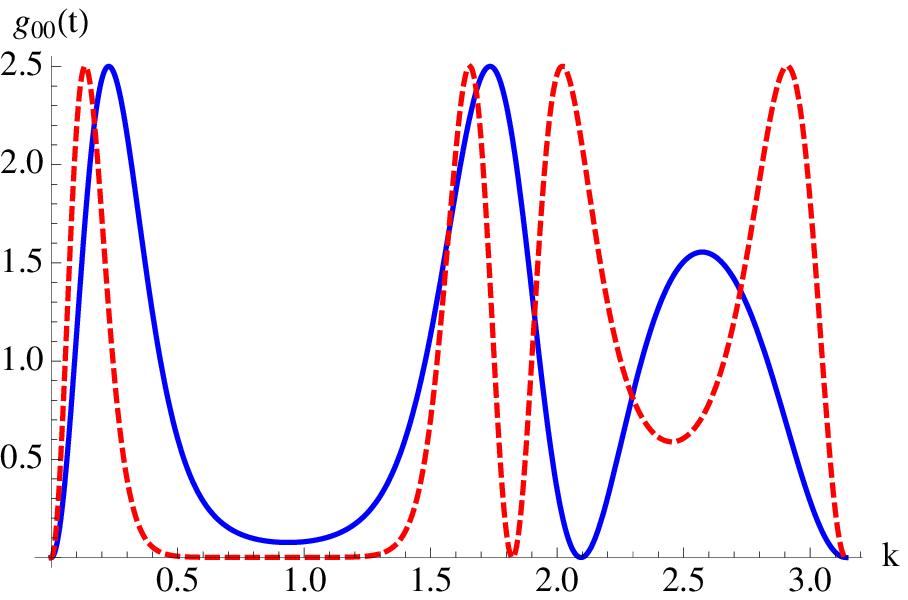}
\caption{$g_{00}(t)$ as a function of $k$ for various values of $J_y$. See text for details.}
\label{g00q1}
\end{minipage}
\hspace{0.4cm}
\begin{minipage}[b]{0.5\linewidth}
\centering
\includegraphics[width=3.0in,height=2.3in]{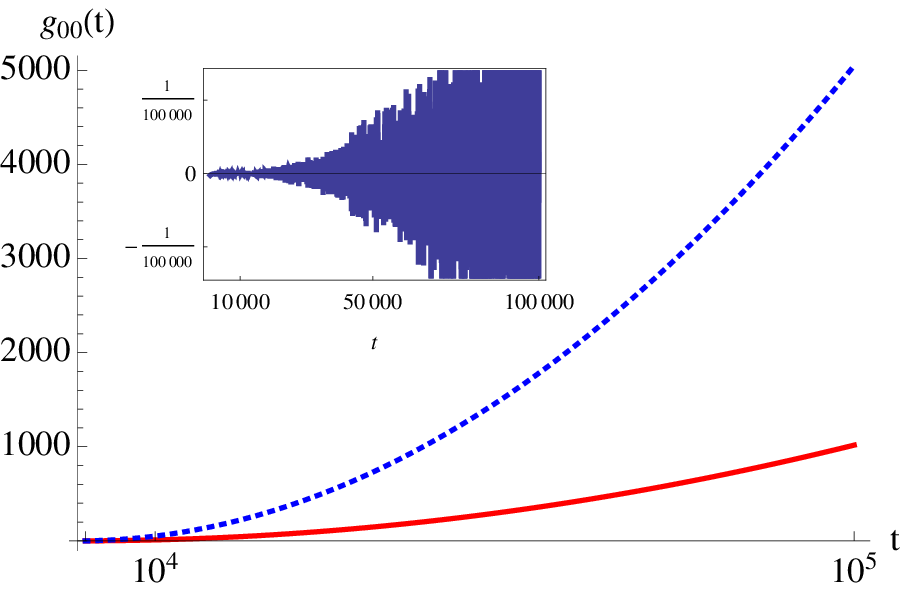}
\caption{$g_{00}(t)$ as a function of $t$ for various values of $k$. See text for details.}
\label{g00q2}
\end{minipage}
\end{figure}

In fig.(\ref{g00q1}), we have shown the behaviour of the metric component $g_{00}(t)$ for $t = 10^3$ (we remember to choose a large time compared to the 
time scale $\tau$ that we have set to unity) as a function of the momentum $k$. The solid blue line corresponds to $J_y=0.5$ and the dashed red line to $J_y=1$. 
In both the cases, we find that $g_{00}(t)$ is positive everywhere except for $k = \cos^{-1}(-h/(2J_y)$ where it reaches zero. This corresponds to the fact that
at this value of the wave vector, there is a quantum phase transition, and the energy gap of eq.(\ref{gap}) closes, indicating non-adiabaticity. 
In fig.(\ref{g00q2}), we have shown the the behaviour of $g_{00}(t)$ for a fixed value of $k= \pi/4$ (solid red) and $k=\pi/6$ (dotted blue) for large times, where, for
both we have chosen $J_y = h = 0.5$. The inset shows the time evolution of this metric component for $k = \cos^{-1}(-0.5)$. We see that $g_{00}(t)$ rapidly
oscillates and time averages to zero. 

There are three important remarks that we make at this stage. First, note that $g_{00}(t)$ should be real, by definition (it is the 
square of the energy uncertainty and is positive definite). For single particle two level systems considered in 
the beginning of this section, it can be checked that $g_{00}$ is manifestly real. 
Here, we find that the imaginary part of $g_{00}(t)$ is not identically zero, but time averages to zero for all values of the wave vector $k$.

Secondly, as a function of $k$, the determinant of the information metric is positive definite everywhere except at $k = \cos^{-1}(-h/2J_y)$ where it is zero at
all times. At this value of $k$, $g_{00}(t)$ is also zero for all times, and this is related to the fact that for this value of the momentum, the probability of finding 
the system (which initially started in the state $|e_{1k}\rangle$) in state $|e_{2k}\rangle$ is unity (see eq.(13) of \cite{Amit}), i.e the system is in a definite 
energy eigenstate and $g_{00}(t)=(\Delta E)^2 =0$. We had encountered such a situation in the two state oscillatory model of the last subsection, where $g_{00}$
vanished in the limit that the coupling constant went to zero, and we saw that in spite of the fact that the metric was seemingly degenerate, the scalar curvature 
there was well defined in the limit. Here, on the other hand, for $k = \cos^{-1}(-h/2J_y)$, we find that the $R$ becomes indeterminate. It can also be checked 
that excepting for this value of $k$, the scalar curvature evaluates to $R=8$ everywhere, indicating that the PM is spherical. 

This last remark deserves attention. Indeed, in our previous examples (spin in a rotating magnetic field or Rabi oscillations), we had a manifestly periodic
system, one which came back to itself after a given time. In the present case, this is not true. So the notion of a spherical PM needs to be explained. In this 
context, we note that the metric on the PM for single spin systems can be written in terms of a periodic coordinate $\phi$ or, equivalently, an ``increasing'' coordinate $t$ (see 
discussion following eq.(\ref{Wangeq})). The important point is that for a periodic PM, some of the components of the metric tensor should show periodic
(i.e oscillatory) behaviour. Here, we find that while $g_{00}$ (when non-zero) is a strictly increasing function of time (see fig.(\ref{g00q2})), $g_{01}$ and $g_{11}$ show 
rapidly oscillatory behaviour with time which does not average to zero for large times. Hence, there is an effective angular variable on the PM due to which
it shows spherical nature. As in the case of Rabi oscillations, it is difficult to find a manifestly spherical form of the metric, but the scalar curvature captures
the essential information about the global structure of the PM, independent of the specific coordinates used. Because of the spherical nature of the manifold,
this system also develops a geometric phase. However, it is difficult to identify the exact time period of oscillations and we will not comment upon this 
issue further. 
 
\subsection{Summary of Section 2}

We now briefly summarize the main results of this section. Here we have considered IG for time dependent systems, focusing on two-level systems. Expectedly,
the geometry of the PM is spherical for single particle systems. For the transverse XY spin chain, we constructed the IG for the Hamiltonian projected on a two level system, 
the situation describing a quantum quench across a critical line. Here, we found that the PM is spherical except for a single value of the wave vector, where a 
quantum phase transition occurs. We contrasted the situation in this latter case with the simpler example of single particle systems. The main conclusion here
is that IG captures the information about phase transition, even in time dependent situations, and is spherical even in non periodic systems. 
 
\section{Classical vs Quantum Holonomy of the Parameter Manifold}

Having elucidated the nature of the PM in some details, we now ask the following 
question : what does the classical holonomy of the PM teach us about the geometric phase of the system. By classical, we mean the angle by which a 
classical vector differs from itself after taken around a closed loop in a curved manifold. This is closely related to, but not equal to the quantum holonomy or the geometric phase, 
which is obtained by solving a time dependent Schrodinger equation. Let us start by illustrating this with the example of the two-sphere. 

\subsection{Classical Holonomy on the Sphere}

Computing the classical holonomy on a two sphere is a textbook problem, and can be found in introductory treatises on 
geometry (see, e.g problem 7.16, Page 45 of \cite{ProblemBookGR}) which we now briefly recapitulate. Here, one starts by parallel transporting a vector along a 
sphere and compares it with the original vector upon coming back to the starting point. Let us say that we parallel transport a vector $V^{\alpha}$,
$\alpha = (\theta, \phi)$, along a small circle on the two-sphere (the polar angle $\theta = {\rm constant}$). The condition for parallel transport means that the covariant
derivative of the vector along the curve is zero, 
which mathematically boils down to $\partial_\phi V^{\alpha} + \Gamma^{\alpha}_{\mu \phi}A^{\mu}=0$, where the 
Christoffel symbol are defined by $2\Gamma^{\alpha}_{\mu\beta} = g^{\alpha\gamma}(g_{\gamma\mu,\beta} + g_{\gamma\beta,\mu} - g_{\mu\beta,\gamma})$.
For the two-sphere, there are only two non-vanishing Christoffel symbols, and inserting them in the geodesic equation, one obtains the
set of equations $\partial_{\phi}V^{\phi} = -\cot\theta~V^{\theta}$ and $\partial_{\phi}V^{\theta}=\sin\theta\cos\theta~V^{\phi}$. This set of 
equations can be solved to obtain $V^{\alpha}$ as a function of the coordinates $(\theta,\phi)$, with appropriate boundary conditions. 
If we initially started with the vector ${\hat \theta}$, i.e $A^{\alpha} = (1,0)^T$, then after parallel transporting by an angle $2\pi$, we obtain
$A^{\alpha} = (\cos(2\pi\cos\theta), - \sin(2\pi\cos\theta)/\sin\theta)^T.$

In order to make the connection with quantum mechanics more apparent, one usually complexifies the basis of unit vectors \cite{Berry}. 
For example, if we started with a normalised initial vector (at $\phi=0$) as $V^{\alpha} = {\hat \theta} + i{\hat\phi}$ $\equiv(1,i/\sin\theta)^T$, then it is not difficult to
see that after a rotation by $2\pi$, the resulting vector is $e^{i\beta_c}(1,i/\sin\theta)^T$, with $\beta_c = 2\pi\cos\theta$, the classical phase angle.
A quantum mechanical computation by solving the time dependent Schrodinger equation on the other hand yields for a spin $1/2$ particle, the 
Berry phase $\beta = -\pi(1\mp \cos\theta)$ with the $\mp$ sign corresponding to the up and down states, respectively. For non-adiabatic evolution,
an entirely similar result holds, with $\theta$ replaced by $\alpha$, in the notation following the discussion of eq.(\ref{Wangeq}). The classical phase 
$\beta_c$ is thus closely related to the Berry phase or the geometric phase. In generic cases therefore, we expect that the classical holonomy of the
PM, determined by the metric of eq.(\ref{metric}) should give us useful information about the geometric phase. 
 
\subsection{Holonomy for theTransverse XY Spin Chain}

We now discuss the classical holonomy of the PM of the transverse XY spin chain encountered in section 2.2, in the thermodynamic limit. Let us first review 
the geometry of the system as seen by its ground state\cite{Polkovnikov}. Geometric phases in the XY spin chain is studied by introducing an additional 
parameter $\phi$, which corresponds to rotation of all the spins simultaneously, about a $z$-axis. This transformation does not affect the energy spectrum of the model, and the 
critical lines are the same as the model discussed in section 2.2. Once the Hamiltonian of eq.(\ref{xyHamil}) is suitably modified \cite{Polkovnikov}, and the 
ground state is obtained, the metric on the PM can be found by standard complex analysis techniques, in the 
thermodynamic limit. It is more convenient here to define a new set of parameters, $J_x=(1+\gamma)J/2$, $J_y=(1-\gamma)J/2$, and set the overall 
scale $J=1$. Then, the coordinates on the three dimensional PM are $(h,\gamma,\phi)$. We first record the expressions for the components of the metric 
tensor (eq.(41) of \cite{Polkovnikov}) in terms of these coordinates :
\begin{equation}
g_{hh} = \frac{1}{16}\frac{1}{|\gamma|\left(1-h^2\right)},~~
g_{\gamma\gamma} = \frac{1}{16}\frac{1}{|\gamma|\left(1+|\gamma|\right)^2},~~
g_{\phi\phi} = \frac{1}{8}\frac{|\gamma|}{\left(1+|\gamma|\right)}
\label{MetricXY}
\end{equation}
For the time being, let us restrict ourselves to the case $\gamma > 0$. To study classical holonomy in the PM defined by the metric
of eq.(\ref{MetricXY}), let us first consider the $h-\phi$
plane. By using a coordinate transformation $h = \cos \alpha$, it is easy to see that the $h-\phi$ plane is flat. In fact, it is cylindrical in shape,
as shown in section 1D of \cite{Polkovnikov}. There is thus no curvature in this plane, and we can thus conclude that the ground state does not pick up 
any geometric phase factor in the $h-\phi$ plane.

The situation is more interesting in the $\gamma - \phi$ plane. The line element, upon a coordinate transformation $\gamma = \tan^2\theta$ becomes 
\begin{equation}
ds^2 =  \frac{1}{16}\frac{d\gamma^2}{\gamma\left(1+\gamma\right)^2} +  \frac{1}{8}\frac{\gamma d\phi^2}{\left(1+\gamma\right)} \equiv
 \frac{1}{4}\left(d\theta^2 + \frac{\sin^2\theta}{2} d\phi^2\right)
\end{equation}
The second equality implies that the metric is somewhat different from that of a two-sphere. To glean more insight into this, we note that for small
values of $\theta$, $ds^2 = d\theta^2 + (\theta/\sqrt{2})^2d\phi^2$, which resembles the standard metric of a conical defect with deficit angle 
$2\pi\left(1-1/\sqrt{2}\right)$.\footnote{This is related to the fact that for small $\gamma$, the Hamiltonian for the model can be cast into a real form, and its eigenvalues
show a conical intersection \cite{Carollo}, \cite{Carollo1}.}
In any case, this is an indication that a singularity develops at tip, where $\theta = \gamma = 0$ (this has also been reported in
the analysis in section 1-D (in particular, figure 3) of \cite{Polkovnikov}). The scalar curvature calculated from this metric also shows a delta function singularity
at $\gamma = 0$. The upshot of the previous discussion is that the classical holonomy in the 
$\gamma-\phi$ plane is now given by $\beta_c = 2\pi\cos\theta/(\sqrt{2})$ (since $\phi$ now runs up to $2\pi/\sqrt{2}$ rather than $2\pi$) 
which equals $2\pi/\sqrt{2(1+\gamma)}$ (since $\gamma=\tan^2\theta$). 
In the general case where $\gamma$ is allowed to be negative, this is more appropriately replaced by $\beta_c = 2\pi/\sqrt{2(1+|\gamma|)}$. This last
fact can be checked by a first principles calculation outlined in subsection 3.1.

To summarise, we see that the classical holonomy for the PM of the transverse XY spin chain indicates that this should be independent of 
the transverse field $h$, in the region $|h|<1$, in the thermodynamic limit. This seems to be an important feature that we expect to carry over to the
geometric phase as well. This last statement is however in contradiction to results appearing in the literature where it has been reported that the
Berry phase does depend on $h$ in the region $|h|<1$ (see figure 1(a) of \cite{Zhu}). 
\begin{figure}[t!]
\centering
\includegraphics[width=3.0in,height=2.3in]{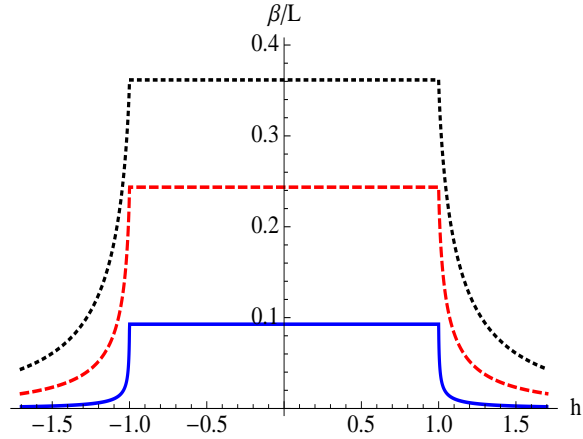}
\caption{Berry Phase for the transverse XY spin chain for different values of $\gamma$. See text for details.}
\label{Berry1}
\end{figure}

We will now seek to resolve this puzzle. First we note that the Berry phase for the XY spin chain was obtained in \cite{Carollo}
\footnote{As we have just mentioned, from a geometric analysis of the PM, we get $0<\phi<\pi/\sqrt{2}$. Whether this modification needs to be made in the definition
of the Berry phase seems to be a subtle issue which we will ignore at the moment.}. The result is \footnote{in \cite{Zhu},\cite{Carollo}, the angle of rotation of the spins about
the $z$-axis is taken to be twice that of \cite{Polkovnikov}. This will not affect the following discussion.} :
\begin{equation}
\beta = -i\int_0^{\pi} \langle g|\partial_{\phi} g \rangle d\phi = \frac{\pi}{N}\sum_{k>0}\left(1 - \cos \theta_k\right)~,
\end{equation}
with $N$ specifying the lattice size, and $\tan\theta_k = \gamma\sin x_k/(h-\cos x_k)$, with $x_k = 2\pi k/L$, $L = 2N + 1$. In the thermodynamic limit $N \to \infty$, one 
can replace $x_k$ by a continuous variable $x$, and the summation by an integration, $\sum_{k=1}^{N} \to L/(2\pi)\int_0^{\pi} dx$. Then, in this limit, the Berry phase is
obtained as 
\begin{equation}
\frac{\beta}{L} = \frac{1}{2}\int_0^{\pi}(1-\cos\theta_k)dx~. 
\end{equation}
Now, from the definition of $\theta_k$, we obtain 
\begin{equation}
\cos\theta_k = \left[\frac{\left(h-\cos x\right)^2}{(h-\cos x)^2 + \gamma^2\sin^2 x}\right]^{1/2}~.
\label{Oureq}
\end{equation}
The expression used in \cite{Zhu} (in which the Berry phase depends non trivially on $h$ in the region $|h|<1$), on the other hand reads 
\begin{equation}
\cos\theta_k = \frac{\cos x-h}{\left[(h-\cos x)^2 + \gamma^2\sin^2 x\right]^{\frac{1}{2}}}~.
\label{Zhueq}
\end{equation}
That these two expressions do not yield the same result upon integration from $0$ to $\pi$ is easily seen (one can simply note that 
$\int_0^{\pi}\sqrt{\cos^2 x}dx\neq \int_0^{\pi}\cos x dx$). Specifically, the natural definition of $\theta_k$ arises via the relation
$\tan\theta_k = \gamma\sin x_k/(h-\cos x_k)$. This allows the range $-\pi/2<\theta_k<\pi/2$, in which $\cos\theta_k$ is always positive. 
However, $\cos\theta_k$ defined from eq.(\ref{Zhueq}) is allowed to be negative, and $\theta_k$ here is in the range $0<\theta_k<\pi$. 
Mathematically, this is the difference between the two definitions and as we show below, $\theta_k$ computed from eq.(\ref{Oureq}) yields
results consistent with the classical holonomy analysis. 

To see this, we use eq.(\ref{Oureq}) to numerically compute the quantity $\beta/L$, for different values of $\gamma$. 
The results of the numerical integration are depicted in fig.(\ref{Berry1}), where the solid blue, dashed red and dotted black lines correspond to the values $\gamma = 0.1$,
$0.3$ and $0.5$, respectively. It is clearly seen that the Berry phase does not depend on $h$ in the range $|h| < 1$.

There are a few issues that need to be addressed here. First of all, the behaviour of the Berry Phase at $\gamma \to 0$ (the XX model limit). In this case, the 
classical holonomy $\beta_c = 2\pi/\sqrt{2}$. The factor of $\sqrt{2}$ indicates that the holonomy is non-zero even for small $\gamma$. In the quantum case, this issue has
been discussed in \cite{Carollo}, \cite{Zhu}. From eq.(\ref{Oureq}), we see that for small $\gamma$, there is always a solution $x = \cos^{-1}(h)$ that
makes $\theta_k = \pi/2$, and hence the Berry phase is non-trivial even in this limit. Also, it is interesting to look at the Berry phase as a function of the anisotropy 
parameter. We find here that as a function of $\gamma$, the derivative $d\beta/d\gamma$ suffers a discontinuity at $\gamma=0$. This might be an important
issue for future investigation.  

Further, our analysis indicates that the derivative $d\beta/dh$ is zero everywhere in $|h|< 1$, in contrast to the result of \cite{Zhu}. In the region $|h| > 1$,
this derivative diverges close to $|h|=1^+$. We should mention here that in the region $|h|>1$, the classical holonomy becomes somewhat
complicated. This is because the parameter manifold is now three dimensional, and from the metric given in eq.(41) of \cite{Polkovnikov}, it can be seen that there 
exists non-trivial classical holonomy both in the $h-\phi$ and the $\gamma-\phi$ planes (in contrast to the case $|h|<1$, where the holonomy in the 
$h-\phi$ plane was trivial). We computed the derivative $d\beta_c/dh$ for both these planes, and 
found that while on the $h-\phi$ plane, $d\beta_c/dh\sim (h-1)^{-1/2}$ for $h\to 1^+$, it becomes a constant in the $\gamma-\phi$ plane. It is difficult to relate
this to the divergence of $d\beta/dh$ for $|h|>1$. 
The exact nature of this divergence, its critical exponent and its relation with the universality of the XY model is an interesting issue and will be addressed in a future work \cite{InPrep}. 

\subsection{A Novel Geometric Phase in the Transverse XY Model}

\begin{figure}[t!]
\centering
\includegraphics[width=3.0in,height=2.3in]{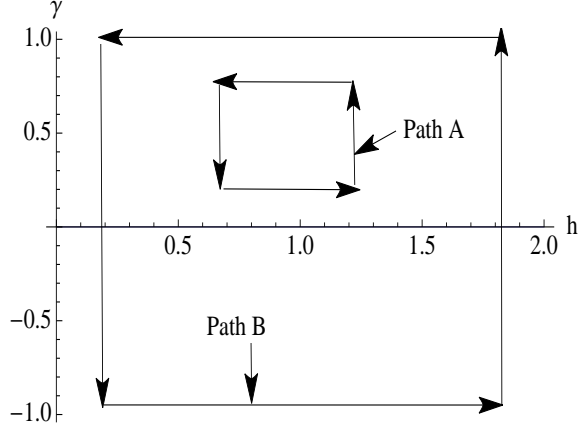}
\caption{Two different closed paths in the $h-\gamma$ plane. See text for details.}
\label{BerryPaths}
\end{figure}

In the previous subsection, we studied the Berry phase associated with the transverse XY model. Here, we briefly comment on the classical holonomy of 
a different type in the same model. Specifically, we compute the holonomy for the model in the $h-\gamma$ plane, as shown in fig.(\ref{BerryPaths}) for two
different paths. The novelty of this computation is that one of the paths cross a critical line, rather than encircle a critical point. 

In the path labeled ``A,'' the coupling constants are varied so that the system never crosses a critical point. For the path labeled ``B,''
the system crosses the critical line at $\gamma = 0$. Importantly, all the path segments are parallel to either the $h$ axis or the $\gamma$ axis. 
Say path ``A'' consists of varying the magnetic field $h$ between $h_1$ and $h_2$ ($h_2 > h_1$, and the anisotropy parameter from $\gamma_1$ and $\gamma_2$,
$\gamma_2 > \gamma_1$. 
In this region, $\gamma >0$ and the line element reads $ds^2 = g_{hh}dh^2 + g_{\gamma\gamma}d\gamma^2$, with $g_{hh}$ and $g_{\gamma\gamma}$
given from eq.(\ref{MetricXY}). It is convenient to first perform a coordinate transformation $h = \sin({\tilde h})$, so that the line element simplifies to
$ds^2 =g_{{\tilde h}{\tilde h}}d{\tilde h}^2 + g_{\gamma\gamma}d\gamma^2$, with $g_{{\tilde h}{\tilde h}}d{\tilde h}=1/(16\gamma)$. Note that $-\pi/2<{\tilde h}<\pi/2$.
Now by a slight abuse of notation (and in order not to clutter the presentation), we will proceed by calling ${\tilde h}$ as $h$. 

Let us first record the expressions for the non-zero Christoffel symbols :
\begin{equation}
\Gamma^h_{\gamma h} = \Gamma^h_{h \gamma}=-\frac{1}{2\gamma},~~\Gamma^{\gamma}_{\gamma \gamma}=-\frac{1+3\gamma}{2\gamma(1+\gamma)},~~
\Gamma^{\gamma}_{hh} = \frac{\left(1+\gamma\right)^2}{2\gamma}~.
\end{equation}
Now, on this PM, we consider parallel transporting a vector with components $(V^h,V^{\phi})^T$, for a fixed $\gamma=\gamma_1 >0$. 
As in the case of the  two sphere discussed in subsection (3.1), this gives rise to the set of coupled equations 
\begin{equation}
\partial_h V^{\gamma} = \frac{\left(1+\gamma_1\right)^2}{2\gamma_1}V^h,~~\partial_h V^{h} = -\frac{1}{2\gamma_2}V^{\gamma}~,
\end{equation}
which can be solved in terms of two unknown coefficients which can in turn be determined from initial conditions. 
The calculation is straightforward, and we will not belabour the details here. We only state the final result that if we had started from a complexified
normalised vector $V_i^{\alpha} = (4/\sqrt{2}(1+\gamma_1)\sqrt{\gamma_1},~4i\sqrt{\gamma_1}/\sqrt{2})^T$, then the vector, after parallel 
transport from $(h_1,\gamma_1)$ to $(h_2,\gamma_1)$ is given by $V_i^{'\alpha} = e^{-i\delta(h_2 - h_1)}V_i^{\alpha}$, with $\delta = (1+\gamma_1)/(2\gamma_1)$.

An entirely similar exercise can be carried out for the sequence $(h_2,\gamma_1)$ $\to$ $(h_2,\gamma_2)$ $\to$ $(h_1,\gamma_2)$ and finally back to
the original point $(h_1,\gamma_1)$ depicted as path ``A'' in fig.(\ref{BerryPaths}). We find that the final vector $V_f^{\alpha}$, after the closed loop is traversed, is now 
related to the initial vector by
\begin{equation}
V_f^{\alpha}= e^{i\beta_c}V_i^{\alpha},~~\beta_c =  \frac{(h_2-h_1)}{2}\left(\frac{1}{\gamma_2} - \frac{1}{\gamma_1}\right)
\label{shrink}
\end{equation}
For path ``B,'' the computation is similar, but needs a little more work. Here, the non-zero Christoffel symbols are given by 
\begin{eqnarray}
\Gamma^h_{\gamma h} &=& \Gamma^h_{h \gamma}=-\frac{\Theta(\gamma)}{|\gamma|} + \frac{1}{2|\gamma|},~~
\Gamma^{\gamma}_{\gamma \gamma}=-\frac{1+3|\gamma|}{|\gamma|(1+|\gamma|)}\Theta(\gamma) + \frac{1+3|\gamma|}{2|\gamma|(1+|\gamma|)}~,\nonumber\\
\Gamma^{\gamma}_{hh} &=& \frac{\left(1+|\gamma|\right)^2}{|\gamma|}\Theta(\gamma) - \frac{\left(1+|\gamma|\right)^2}{2|\gamma|}~.
\end{eqnarray}
where $\Theta(\gamma)$ is the step function which is unity for $\gamma >0$ and equals zero for $\gamma < 0$. We compute the classical holonomy around a
path $(h_1,-\gamma_1)$ $\to$ $(h_2,-\gamma_1)$ $\to$ $(h_2,\gamma_2)$ $\to$ $(h_1,\gamma_2)$ and back to $(h_1,-\gamma_1)$ depicted as 
path ``B'' in fig.(\ref{BerryPaths}). Again, denoting a normalised initial vector by $V_i^{\alpha}$, a straightforward but somewhat lengthy computation yields the final 
vector and hence the classical phase as 
\begin{equation}
V_f^{\alpha}= e^{i\beta_c}V_i^{\alpha},~~ \beta_c = \frac{\left(h_2-h_1\right)}{2}\left(2 + \frac{1}{\gamma_2} + \frac{1}{\gamma_1}\right)~.
\label{noshrink}
\end{equation}
A few comments about the two results of eq.(\ref{shrink}) and (\ref{noshrink}) are in order. In the former, if we set $h_1 \to h_2$, i.e 
$h_2 - h_1 = \Delta$, where $\Delta$ is a small number, and further set $\gamma_1 - \gamma_2 \sim {\mathcal O}(\Delta)$, the classical
phase $\beta_c \sim {\mathcal O}(\Delta^2)$. In the latter case, setting $\gamma_1 \to -\gamma_2$ as appropriate (we took one of the values of
$\gamma$ to be negative), we obtain $\beta_c \sim {\mathcal O}(\Delta)$. This points to the difference between the classical phases as the
curves are shrunk to small size, in the case where the loop does not enclose critical points and in the case that it does. 

We however note that there is a qualitative difference between this case and the one considered in \cite{Hamma}. In that paper, it was shown that
the Berry phase on a non-contractible curve (i.e one that encircles a quantum critical point) is an indicator of quantum phase transitions. Here what we
have considered is a curve that on the other hand passes through a critical line. The discussion in the previous paragraph indicates that in this 
case, the classical phase can be made arbitrarily small, and one can expect a similar result in the geometric phase as well. 

Admittedly, the above was a computation of the classical holonomy of the transverse XY spin chain in the $h-\gamma$ plane. However, we believe
that as in the examples considered in this section, the broad features of this analysis should be related to the geometric phase when one considers an appropriate 
time dependent Schrodinger equation. In the quantum case, such a situation would involve non-adiabatic evolution as the system is driven through a 
critical line. It would be interesting to study this further. 

\subsection{Summary of Section 3}

We will now briefly summarise the main results of this section. We set out by asking what we can learn about the Berry phase (and in general the geometric phase)
from the classical holonomy calculation on the PM of many body quantum systems. This is important, as in general, the classical computation can yield 
analytic results for sufficiently simple PMs. We took the example of the transverse XY spin chain model, and found some interesting predictions from the classical
viewpoint, which we confirmed by a first principles calculation from the quantum many body ground state. We emphasise that our results are different from the
ones that have appeared on the topic before \cite{Zhu}, but believe that we have given enough evidence about the correctness of our computations. The
issue certainly deserves further analysis. 
Further, we considered a loop in the PM of the XY spin chain which cannot be described by an adiabatic time evolution of the ground state, and calculated its 
classical holonomy. We are tempted to speculate that this is related to a quantum geometric phase of the model. 

\section{Discussions and Conclusions}

In this paper, we have studied two issues related to the information geometry of time dependent quantum systems. For two dimensional
PMs (with time and one coupling constant as the coordinates) the PM is spherical in nature for two level single particle systems, as expected. 
Importantly, we showed that an exception arises in the case of an anisotropic quantum quench in the 
transverse XY spin chain studied in subsection 2.2. In that case, we found that the scalar curvature of the PM is a positive constant everywhere except for
the quantum critical point. This indicates that IG correctly captures the behaviour of time evolving many body systems. 

The second issue that we commented upon was the role of classical holonomy for PMs corresponding to quantum many body systems in the thermodynamic limit. 
We found that as in single spin examples, this captures important information regarding the geometric phase. Using this, we argued that for the transverse XY model, the
available literature on the Berry phase needs a re-look. Further, we computed a classical 
phase in the transverse XY model that involves non-adiabatic evolution, and conjectured that this might be related to the quantum geometric phase 
in the model. It might be useful to extend this analysis further. 

\vspace{0.2in}
\begin{center}
{\bf Acknowledgements}
\end{center}
It is a pleasure to thank V. Subrahmanyam for fruitful discussions.

\end{document}